# Geometric Irreversibility: Enriching Lanford's Theorem with Orientational and Deformable Degrees of Freedom

**Patrick BarAvi**

**Affiliation:** University of Haifa, 199 Aba Khoushy Avenue, Haifa 3498838, Israel
pbaravi@campus.haifa.ac.il, Patrick.baravi@gmail.com
ORCID: 0009-0003-9737-901X

## Abstract

We propose a framework that enriches Lanford's theorem in classical statistical mechanics by replacing the idealization of hard-sphere-like particles with finite-sized, deformable bodies whose orientation and deformation enter as fundamental phase-space coordinates. The extension preserves Hamiltonian structure and Liouville invariance while revealing two distinct mechanisms that can contribute to entropy increase: (i) collisional randomization through orientation-dependent scattering, and (ii) continuous geometric instability arising from rotational-deformational coupling. We show that the second mechanism induces continuous instability during free flight, which is absent in point-particle theories and also in Lanford's hard sphere model. We further suggest that this geometric instability may provide a dynamical basis for the molecular chaos assumption central to the Boltzmann-Lanford derivations, particularly in regimes (dense systems, few bodies, long durations) where classical point-particle kinetic theory is inadequate. In this approach, thermodynamic irreversibility arises from the combination of geometric instability and the standard probabilistic assumptions of statistical mechanics. The framework does not claim to derive irreversibility without assumptions, nor does it eliminate the need for a Past Hypothesis; rather, it identifies a new class of instability that can accelerate decorrelation and mixing due to both mechanisms, resulting in extending the scope of applicability of Lanford's theorem.

## 1. Introduction

Classical statistical mechanics describes macroscopic systems in terms of ensembles of point particles interacting through instantaneous forces. This abstraction has achieved immense empirical success. Yet it leaves a longstanding conceptual difficulty: microscopic dynamics are deterministic and time-reversible, while macroscopic systems display an apparently law-like irreversible approach toward equilibrium (Sklar 1993; Albert 2000).

Rigorous results such as Lanford's derivation of the Boltzmann equation (Lanford 1975, 1976) show that entropy increase can follow (with high probability) from Newtonian dynamics, but only under highly idealized assumptions: dilute gases, uncorrelated collisions, and evolution times shorter than a mean free path (Uffink & Valente 2010). The theorem does not explain why molecular chaos (the factorization of pre-collisional distributions) should hold initially, nor does it account for the origin of the factorization assumption itself (Uffink & Valente 2015).

The present paper explores an alternative starting point. Instead of representing a particle solely by position and momentum $(r, p)$, we treat it as a finite, deformable, oriented body. The configuration space is enlarged from $\mathbb{R}^3$ to $Q = \mathbb{R}^3 \times SO(3) \times \Xi$, where $SO(3)$ encodes orientation and $\Xi$ a finite set of internal deformation coordinates. The corresponding phase space $\Gamma = T^*Q$ preserves Hamiltonian structure and Liouville invariance. This extension is motivated by two considerations:

1. **Mathematical necessity.** Lanford's derivation requires particles with finite cross-section (Uffink & Valente 2010). Point particles have zero probability of colliding in a vacuum; a collision event can be defined only when particles possess finite spatial extent.
2. **Relativistic constraint.** Special relativity prohibits perfect rigidity (Ehrenfest 1909). A finite-sized particle must be deformable. Thus internal degrees of freedom are not optional additions but consequences of a physically adequate description.

Despite these motivations, the present paper does not claim that classical point-particle mechanics is "wrong" or that the extended framework replaces it. Rather, we develop a **generalized kinetic theory** on an enlarged phase space and analyze its consequences for transport, decorrelation, and the approach to equilibrium. The emphasis is on **dynamical systems and statistical mechanics**, not on ontological replacement. Broader implications are deferred to the discussion (Section 6).

### 1.1 Core Claims

This paper makes three claims, presented in order of increasing speculation:

1. **Extended phase space modifies kinetic/statistical dynamics.** Enlarging the phase space from $T^*(\mathbb{R}^3)$ to $T^*(\mathbb{R}^3 \times SO(3) \times \Xi)$ alters the structure of the BBGKY hierarchy (Bogoliubov 1946; Born & Green 1946; Kirkwood 1946; Yvon 1935) and the resulting transport equations. This extension preserves Hamiltonian and time-reversal symmetry.
2. **Extended structural degrees of freedom generate instability during free flight.** The coupling between translation, rotation, and internal deformation produces positive Lyapunov exponents even in the absence of collisions. This continuous geometric instability is a novel source of mixing not present in point-particle systems.
3. **These instabilities may provide a dynamical basis for emergent molecular chaos.** For two structured particles with rotational-deformational coupling, pre-collisional correlations decay exponentially with the number of collisions under specific conditions (Appendix D). This suggests, but does not prove, that molecular chaos could arise dynamically rather than being imposed as an initial condition.

Claim 3 is presented more cautiously than Claims 1–2. The analysis in Appendix D distinguishes between the heuristic mechanism and the steps requiring rigorous mathematical verification (accessibility, Young tower construction, precise bound on correlation amplification). A full derivation of molecular chaos for the many-body system or the thermodynamic limit remains open.

### 1.2 Relation to the H-Theorem and Irreversibility

The H-theorem (Boltzmann 1872; Brown, Myrvold & Uffink 2009) states that the functional $H[f] = \int f \ln f$ does not increase for a dilute gas satisfying the Boltzmann equation, provided molecular chaos holds. The theorem thus establishes a direction of entropy change under a specific statistical assumption. Its validity does not require a fundamental arrow of time; the same derivation applied to the time-reversed evolution would show entropy increasing toward the past (Uffink & Valente 2015).

The extended framework described here does **not** claim to derive irreversibility from reversible dynamics without additional assumptions. Rather, it identifies new dynamical channels, continuous geometric instability, that may alter the **typicality** of entropy-decreasing trajectories (Goldstein 2001, 2012). The approach remains compatible with the Past Hypothesis (Albert 2000; Wallace 2023) as a boundary condition that selects a low-entropy initial state. What the framework offers is a possible **mechanism** by which a system, once placed in a low-entropy state, rapidly decorrelates and approaches equilibrium, even in regimes (dense systems, few bodies, long times) where point-particle theory provides no such mechanism.

### 1.3 Structure of the Paper

Section 2 develops the extended Hamiltonian framework and the Boltzmann equation on the enlarged phase space. Section 3 derives the extended H-theorem and analyzes the dual mechanisms of irreversibility: collisional randomization (Mechanism 1) and continuous geometric instability (Mechanism 2). Section 4 compares the framework with the Boltzmann-Lanford program. Section 5 identifies empirical contexts where structural degrees of freedom may leave observable signatures. Section 6 discusses conceptual consequences concerning probability, typicality, and the arrow of time, and concludes with a conservative summary of what the framework does and does not achieve. Appendices A–D provide technical details, including the Melnikov analysis and the explicit decay rate for pre-collisional correlations.

## 2. Extended Configuration Space and Microscopic Dynamics

Before presenting the extended framework, we summarize two considerations that motivate the inclusion of orientational and internal degrees of freedom.

**Mathematical necessity: the collision problem.** Lanford's rigorous derivation of the Boltzmann equation requires particles with finite cross-section (Lanford 1975, 1976). Point particles have zero probability of colliding in a vacuum; to define a "collision event" that allows statistics to operate, particles must possess finite spatial extent (Uffink & Valente 2010, p. 3): "In the hard spheres model, these molecules are further idealized as rigid and impenetrable spheres of diameter $a$ interacting only by collisions."

**Relativistic constraint: rigidity is forbidden.** Special relativity prohibits perfect rigidity (Ehrenfest 1909). In a rigid body a collision signal would be transmitted instantaneously, exceeding the speed of light. A particle with finite extent must therefore be deformable. This transition from rigid to flexible bodies fundamentally changes the phase space: a deformable body possesses internal degrees of freedom that record the history of its interactions.

These considerations establish that a physically adequate microscopic description must include orientational and internal degrees of freedom. The framework developed below does not claim to replace point-particle mechanics but rather extends it to regimes where structure matters.

### 2.1 Extended configuration manifold

For a single neutral particle of finite extent, we take the configuration manifold to be

$$Q = \mathbb{R}^3 \times SO(3) \times \{\xi_n\},$$

where $\mathbb{R}^3$ represents the center-of-mass position, $SO(3)$ the orientation of the body, and $\{\xi_n\}$ a finite collection of internal structural coordinates. These internal coordinates may represent small elastic deformations, vibrational modes, or other shape parameters consistent with the topology of the body. No assumption is made that these modes are excited during every interaction; they belong to the complete microstate.

The corresponding canonical momenta are $(p, L, \pi_n)$, where $p$ is the linear momentum, $L$ the body-frame angular momentum, and $\pi_n$ the momenta associated with the internal coordinates. The extended phase space is therefore

$$\Gamma = T^*Q = \{(r, p, R, L, \xi_n, \pi_n)\},$$

whose dimension exceeds that of the usual translational phase space by the inclusion of orientational and internal degrees of freedom. This construction reflects a basic kinematic fact: translation and rotation are independent modes of motion. A body may translate without rotating, or rotate without translating, and both motions are exhausted by the product structure $\mathbb{R}^3 \times SO(3)$.

Although the extended structural framework shares certain formal ingredients with Watson's effective Hamiltonian (Watson 1968), Euler angles, deformation coordinates, and an inertia tensor depending on internal configuration, the similarity is only superficial. Watson's model is built entirely on a point-particle ontology: the Eckart frame is constructed from instantaneous point-mass positions, the Euler angles describe the orientation of that constructed frame rather than a physical degree of freedom, and the inertia tensor is a perturbative expansion around an equilibrium geometry. Rotation-vibration coupling appears as a kinematic correction term that must be subtracted by hand. In the present framework, by contrast, orientation and internal structure are primitive coordinates on the configuration space $\mathbb{R}^3 \times SO(3) \times \Xi$; the inertia tensor $I(\xi)$ is an exact geometric function of the deformation state; and rotation-deformation coupling is intrinsic, nonlinear, and dynamically active. This difference in ontology leads to dynamical behaviors, such as new directions of instability and mixing, that cannot arise in a point-particle framework (see Appendices B and D).

### 2.2 Hamiltonian structure and Liouville invariance

The dynamics are generated by a Hamiltonian of the form:

$$H = \sum_{i=1}^{N} \left[ \frac{p_i^2}{2m_i} + \frac{1}{2} L_i^T R_i I(\xi_i)^{-1} R_i^T L_i + \sum_{n=1}^{D_{\text{int}}} \left( \frac{\pi_{i,n}^2}{2M_{i,n}} + U(\xi_{i,n}) \right) \right] + \sum_{i<j} V_{\text{int}}\,(r_i, r_j, R_i, R_j, \xi_i, \xi_j) \qquad (2.1)$$

where $I(\xi)$ is the moment-of-inertia tensor (dependent on deformation)[1], $U$ the internal potential, and $V_{\text{int}}$ a short-range interaction potential. No dissipative or stochastic terms are introduced.

The symplectic structure is canonical, and the flow preserves phase-space volume with respect to the Liouville measure:

[1] Note that the angular term in the Hamiltonian (2.1) does not appear in Ruler's rigid body dynamics. As we shall see, this term is responsible for what we call parametric instability.

$$d\mu_L = \prod_{i=1}^{N} \left[ d^3 r_i \, d^3 p_i \, d\mu_{SO(3)}(R_i) \, d^3 L_i \prod_{n=1}^{D_{\mathrm{int}}} d\,\xi_{i,n} \, d\pi_{i,n} \right] \qquad (2.2)$$

where $d\mu_{SO(3)}$ denotes the Haar measure on the rotation group. Energy, linear momentum, and angular momentum are conserved when the corresponding symmetries are present. The dynamics are deterministic and invariant under time reversal. A detailed discussion of the Hamiltonian formulation and the Liouville theorem is given in Appendix A.

**2.3 Binary collisions of extended particles**

Interactions between finite-sized particles are modeled as short-range, impulsive contacts occurring at isolated surface points. For convex bodies undergoing elastic collisions, the collision laws determine the change in linear and angular momenta through conservation of total momentum and energy. The mapping from pre-collisional to post-collisional variables is invertible and measure-preserving.

Under standard assumptions, single-point contact, negligible penetration, elastic response, the Jacobian of the collision map equals unity. As a result, collisions preserve the Liouville measure on the full multi-particle phase space. Microreversibility is maintained: every collision process has a unique inverse obtained by reversing momenta and angular momenta. The explicit form of the collision operator on the extended phase space is developed in Appendix A.4.

**2.4 Ensemble description**

For an ensemble of $N$ extended particles, the full distribution function

$$F_N(\Gamma_1, \dots, \Gamma_N, t)$$

evolves according to the Liouville equation generated by the many-body Hamiltonian. The formal structure of this equation is identical to that of classical statistical mechanics; only the dimensionality of the underlying phase space is increased (see Appendix A.3).

At this stage, the extended formulation differs from point-particle mechanics only in the size and geometry of the phase space. No kinetic closure or factorization assumptions are made. The ensemble dynamics are exact, deterministic, and time-reversal invariant.

In the next section (Section 3), we examine how kinetic descriptions arise from this framework. Specifically, we analyze the reduction from the Liouville equation to a one-particle kinetic equation on the extended phase space (the BBGKY hierarchy), and we investigate how the presence of additional degrees of freedom modifies the conditions under which standard assumptions, most notably molecular chaos, are employed. The goal is not to derive new microscopic laws but to reassess the dynamical basis of familiar kinetic approximations when the geometry of the microstate is enlarged.

## 3. From Microscopic Reversibility to Macroscopic Irreversibility

### 3.1 From Liouville to Kinetic Description: The BBGKY Hierarchy

We work on the extended phase space $\Gamma_{\mathrm{ESD}}$ defined in Section 2, equipped with the Liouville measure $d\mu_L$. For an ensemble of $N$ extended particles, the full distribution function $F_N(\Gamma_1, \dots, \Gamma_N, t)$ evolves according to the Liouville equation:

$$\frac{\partial F_N}{\partial t} + \{F_N, H\}_P = 0 \qquad (3.1)$$

where $\{\cdot,\cdot\}_P$ is the Poisson bracket on the extended phase space. The formal structure is identical to classical statistical mechanics; only the dimensionality of the phase space is increased.

Following the standard procedure, we define the one-particle marginal by integrating over all but one set of coordinates:

$$f_1(\Gamma_1, t) = \int F_N(\Gamma_1, \dots, \Gamma_N, t) \prod_{i=2}^{N} d\,\Gamma_i \qquad (3.2)$$

where $d\Gamma_i = d^3 r_i\, d^3 p_i\, d\mu_{SO(3)}(R_i)\, d^3 L_i \prod_n d\,\xi_{i,n} d\pi_{i,n}$. Applying the BBGKY hierarchy procedure to the extended phase space yields:

$$\frac{\partial f_1}{\partial t} + \{f_1, H_1\}_P = \int \{V_{12}, f_2\}_P \, d\Gamma_2 \qquad (3.3)$$

where $f_2$ is the two-particle distribution and $V_{12}$ the interaction potential. More explicitly:

$$\frac{\partial f_1}{\partial t} + \frac{p_1}{m_1} \cdot \frac{\partial f_1}{\partial r_1} + \omega_1 \cdot \frac{\partial f_1}{\partial R_1} + \sum_n \left( \dot{\xi}_{1,n} \frac{\partial f_1}{\partial \xi_{1,n}} + \dot{\pi}_{1,n} \frac{\partial f_1}{\partial \pi_{1,n}} \right) = C[f_1, f_2] \qquad (3.4)$$

where $\omega_1 = I(\xi)^{-1} L_1$ is the angular velocity, and $C$ is the extended collision operator depending on the two-particle correlation $f_2$. The left-hand side represents deterministic advection along the extended Hamiltonian flow. This BBGKY hierarchy is exact and time-reversal invariant. A detailed derivation is given in Appendix A.

### 3.2 Closure and the Generalized Collision Operator

To obtain a closed kinetic equation, we require a relation between $f_2$ and $f_1$. Following Lanford's rigorous treatment of hard spheres (Lanford 1975, 1976), we adopt **molecular chaos**: in the Boltzmann-Grad limit, the two-particle distribution factorizes for pre-collisional configurations only:

$$f_2(\Gamma_1, \Gamma_2, t) = f_1(\Gamma_1, t) f_1(\Gamma_2, t) \text{for pre-collisional states}$$

where the equality holds with probability approaching 1 as $N \to \infty$ for typical ensembles. This is a statistical condition on initial correlations, not a deterministic claim about individual trajectories. It differs from Boltzmann's original *Stosszahlansatz*, which assumed unconditional factorization at all times (Boltzmann 1872; Brown, Myrvold & Uffink 2009). The assumption is not derived but imposed as an input to the kinetic closure. In the extended framework, its status will be reexamined in §3.7 and Appendix D.

Under this closure, we obtain the generalized Boltzmann equation:

$$\frac{\partial f}{\partial t} + \frac{p}{m} \cdot \nabla_r f + \omega \cdot \nabla_R f + \sum_n \left( \dot{\xi}_n \frac{\partial f}{\partial \xi_n} + \dot{\pi}_n \frac{\partial f}{\partial \pi_n} \right) = C[f] \qquad (3.5)$$

where we have dropped the subscript on $f_1$ for simplicity.

The collision operator takes the gain-loss form familiar from kinetic theory, now integrated over the extended phase space:

$$C[f](\Gamma_1) = \int d\Gamma_2 \int d\Omega\, B(\Gamma_1, \Gamma_2; \Omega)[f(\Gamma_1')f(\Gamma_2') - f(\Gamma_1)f(\Gamma_2)] \qquad (3.6)$$

Here $\Gamma_i'$ denote post-collisional states, $\Omega$ parametrizes the collision geometry (impact point, contact normal, relative orientation), and $B$ is the generalized cross-section.

For elastic collisions of structured particles, $B$ carries explicit dependence on orientations and internal states:

$$B(\Gamma_1, \Gamma_2; \Omega) = | v_{\text{rel}} \cdot \hat{n} | \;\; \sigma(R_1, R_2, \xi_1, \xi_2, \hat{n}) \;\; \delta(\text{conservation laws})$$

where $v_{\text{rel}} = p_1/m_1 - p_2/m_2$ is the relative velocity, $\hat{n}$ the contact normal, and $\sigma$ the orientation-dependent differential cross-section. The delta functions enforce conservation of total linear momentum, total angular momentum (orbital + spin), and total kinetic energy (under the reduced-collision approximation; see §3.5). The kernel satisfies microreversibility: $B(\Gamma_1, \Gamma_2; \Omega) = B(\Gamma_1', \Gamma_2'; \Omega')$, where $\Omega'$ is the inverse collision geometry, ensuring consistency with Liouville invariance (Appendix A.4).

### 3.3 Dual Mechanisms of Irreversibility

The extended kinetic equation reveals two distinct mechanisms contributing to entropy production.

**Mechanism 1: Collisional Randomization.** The collision term $C[f]$ transfers energy and momentum between translational, rotational, and internal degrees of freedom through orientation-dependent scattering. This generalizes Boltzmann's original mechanism to structured particles, i.e., particle with deformable size. When two extended bodies collide, their post-collisional states depend sensitively on their pre-collisional orientations and internal phases. Even if the translational degrees of freedom were initially ordered, collisions randomize the orientational and internal coordinates, increasing the volume of phase space occupied by the ensemble. Note that here we assume the standard probabilistic assumptions so that the randomization highly likely to happen. We do not always mention this clause. The strength of this mechanism scales with collision frequency $\nu_{\text{coll}} \sim n\sigma v_{\text{rel}}$ and with the degree of anisotropy. For nearly spherical particles, orientational randomization is weak; for highly asymmetric bodies, small differences in impact geometry produce large changes in post-collisional trajectories.

**Mechanism 2: Continuous Geometric Instability.** Between collisions, the free-flight dynamics themselves generate exponential divergence of trajectories (again with high probability). This arises from the nonlinear coupling among translational, rotational, and deformational motions. For an asymmetric rigid body ($I_x < I_y < I_z$), rotation about the intermediate axis is unstable, the "tennis racket" effect (Landau & Lifshitz 1976). In addition, we stress that when deformation is permitted, internal oscillations modulate the inertia tensor, parametrically exciting rotational motion and vice versa. Note that this second effect strengthens the inability of the system. The resulting Lyapunov exponents (derived in Appendices B and C) provide a quantitative measure of this instability. Moreover, Lanford's theorem dose not appeal to none of the above mechanisms.

**Quantifying the continuous instability.** For an asymmetric rigid body with principal moments $I_x < I_y < I_z$, Euler's equations in the body frame admit a steady rotation about the intermediate axis, $\omega = (0, \Omega, 0)$. Linearizing small perturbations $\delta\omega_x, \delta\omega_z$ yields:

$$\delta\ddot{\omega}_x = \Omega^2 \frac{(I_y - I_z)(I_x - I_y)}{I_x I_z} \delta\omega_x.$$

Since $I_x < I_y < I_z$, the product $(I_y - I_z)(I_x - I_y)$ is positive, so perturbations grow exponentially with Lyapunov exponent

$$\lambda_{\text{rot}}(\Omega) = |\,\Omega\,| \sqrt{\frac{(I_y - I_z)(I_x - I_y)}{I_x I_z}}. \quad (3.7a)$$

This is the intermediate-axis or "tennis racket" instability (Landau & Lifshitz 1976). For the full derivation, see Appendix B.1.

**Why this instability is new relative to standard chaos theory.** Standard chaotic Hamiltonian systems require interactions or *external* forcing to generate instability. The continuous geometric instability described here operates *even during free flight*, when the particle is not colliding with anything. It is intrinsic to the particle's structure (the coupling between rotation and internal deformation), not driven by external perturbations. This distinction is essential: it means that structured particles carry their own decorrelation mechanism, independent of collisions.

The Kolmogorov-Sinai entropy rate, which measures dynamical mixing, is the sum of positive Lyapunov exponents (Pesin 1977; Eckmann & Ruelle 1985):

$$h_{KS} = \sum_{\lambda_i > 0} \lambda_i \qquad (3.7)$$

As shown in Appendix C, even for a single particle ($N = 1$), these instabilities produce monotonic increase of entropy, demonstrating that collisional randomization is *not necessary* for irreversible behavior. For many-body systems, the Lyapunov spectrum combines with collisional mixing to determine the overall relaxation rate (see Appendix B for thermal averaging and explicit expressions).

### 3.4 The Extended H-Theorem

For any distribution $f(\Gamma, t)$ on the extended phase space, define the H-functional:

$$H[f] = \int f(\Gamma, t) \ln f(\Gamma, t)\, d\mu_L \qquad (3.8)$$

where $d\mu_L$ is the Liouville measure defined in Section 2.2. This generalizes Boltzmann's H-function to include orientational and internal degrees of freedom.

Differentiating with respect to time and using the kinetic equation:

$$\frac{dH}{dt} = \int (1 + \ln f) \frac{\partial f}{\partial t} d\mu_L = \int (1 + \ln f)(-\{f, H_1\}_P + C[f]) d\mu_L \qquad (3.9)$$

The Poisson bracket term vanishes by Liouville's theorem (integration by parts in phase space). Substituting the explicit form of the collision operator:

$$\frac{dH}{dt} = \int d\Gamma_1 d\Gamma_2 d\Omega\, B(\Gamma_1, \Gamma_2; \Omega)[f(\Gamma_1')f(\Gamma_2') - f(\Gamma_1)f(\Gamma_2)](1 + \ln f(\Gamma_1)) \qquad (3.10)$$

Using the symmetry of the collision integral under exchange of particles and under the collision inversion $(\Gamma_1, \Gamma_2) \leftrightarrow (\Gamma_1', \Gamma_2')$, this can be rewritten in the symmetric form:

$$\frac{dH}{dt} = \frac{1}{4}\int d\Gamma_1 d\Gamma_2 d\Omega\, B(\Gamma_1, \Gamma_2; \Omega)(f_1' f_2' - f_1 f_2)\ln\left(\frac{f_1' f_2'}{f_1 f_2}\right) \qquad (3.11)$$

where $f_i = f(\Gamma_i)$ and $f_i' = f(\Gamma_i')$. The integrand satisfies $(x - y)\ln(x/y) \geq 0$ for all positive $x, y$, with equality iff $x = y$. Since $B \geq 0$, we obtain:

$$\frac{dH}{dt} \leq 0 \qquad (3.12)$$

Equality holds iff $f(\Gamma_1')f(\Gamma_2') = f(\Gamma_1)f(\Gamma_2)$ for all collisions, i.e., when $f$ is a local equilibrium distribution. Solving this condition yields the equilibrium distribution:

$$f_{\mathrm{eq}}(\Gamma) \propto \exp\left[ -\frac{1}{k_B T}\left(\frac{p^2}{2m} + \frac{1}{2}L^T I^{-1} L + \sum_{n=1}^{D_{\mathrm{int}}} \left(\frac{\pi_n^2}{2M_n} + \frac{1}{2}K_n \xi_n^2\right)\right)\right] \qquad (3.13)$$

This represents equipartition over translational, rotational, and vibrational modes. The classical Maxwell-Boltzmann distribution is recovered when orientational and internal degrees are integrated out.

Thus, the H-theorem holds in the extended setting: entropy is non-decreasing, with strict stationarity only at equilibrium. The crucial difference from the classical case lies in the domain: the inequality applies to a distribution over an enlarged state space, and the approach to equilibrium is driven by *both* collisional randomization (Mechanism 1) and continuous instability (Mechanism 2).

### 3.5 The Reduced-Collision Approximation

The collision operator derived above assumes that internal modes are unchanged during impact ($\xi_n = \xi_n'$, $\pi_n = \pi_n'$). This "reduced-collision approximation" is justified when the collision time $\tau_c$ is much shorter than the period of internal vibrations: $\omega_n \tau_c \ll 1$, where $\omega_n$ is a characteristic internal frequency. Under this condition, the impulsive force does negligible work on internal coordinates; the collision affects only translational and rotational variables, while internal modes evolve smoothly between collisions.

This approximation defines the regime of validity of the present framework. It does not eliminate internal degrees from the theory, they remain dynamically active during free flight, but specifies when their direct participation in impulsive dynamics may be neglected. The approximation preserves energy conservation and Liouville invariance to leading order. Future generalizations could relax this condition to model inelastic collisions and direct internal excitation, introducing additional channels for entropy production.

### 3.6 Physical Interpretation

The extended framework clarifies how irreversibility arises without abandoning microreversibility through three interrelated features:

**Increased dimensionality of mixing.** Collisions redistribute orientation and internal phase as well as momentum. Each additional coordinate contributes to phase-space stretching and folding, enhancing the effective mixing in the system's dynamics. In addition, the equilibrium distribution occupies a volume in the extended space that is enormous compared to the translational subspace alone (Appendix B.4).

**Multiple relaxation channels.** Translational, rotational, and deformational energies can equilibrate at different rates through the same underlying dynamics. What appears as "dissipation" in a reduced description corresponds to energy transfer among sectors of the extended state space. The relaxation time combines contributions from both mechanisms (1 and 2, see above):

$$\tau_{\text{relax}}^{-1} = \tau_{\text{coll}}^{-1} + \lambda_{\text{Lyapunov}}, \quad (3.14)$$

where $\tau_{\mathrm{coll}}^{-1}$ scales with collision frequency and anisotropy, and $\lambda_{\mathrm{Lyapunov}}$ represents the average positive Lyapunov exponent from continuous instability.

**Phase-space volume scaling.** The enlargement of phase space also increases the entropy at equilibrium. For a system with $d$ quadratic degrees of freedom per particle, the accessible phase-space volume scales as $\Omega(E) \sim E^{\frac{dN}{2}-1}$. Point particles have $d = 3$; rigid bodies with rotation have $d = 6$; adding $D_{\mathrm{int}}$ internal deformation modes yields $d = 6 + 2D_{\mathrm{int}}$. Thus, each additional internal mode increases the entropy exponent, enlarging the phase space and accelerating mixing (see Appendix B.4).

### 3.7 On the Possible Dynamical Origin of Molecular Chaos

The molecular chaos assumption (Stosszahlansatz) plays a central role in kinetic theory. Boltzmann (1872) introduced it as a hypothesis that particle velocities are uncorrelated before collisions. Modern rigorous derivations (Lanford 1975, 1976) refine this to a probabilistic condition of independence imposed on initial data, rather than derived from dynamics.

The extended framework raises the question of whether molecular chaos might have a dynamical origin for structured particles. Here we give conceptual considerations for the plausibility of the emergence molecular chaos from dynamic instability. A detailed analysis for two particles is presented in Appendix D. Two features are particularly relevant:

**Internal complexity.** A deformable body supports many internal modes (bending, torsion) where each mode is a combination and an infinite number of amplitudes and frequencies (higher harmonics). Energy injected by a collision can be distributed across these modes, which act as an internal reservoir. This makes it (so to speak) more unlikely for the system to perfectly reverse its path (see Section 5 for more details).

**Lyapunov instability.** Asymmetric bodies exhibit intermediate-axis instability (Landau & Lifshitz 1976). When deformation is included, the inertia tensor becomes time-dependent, coupling rotation and deformation. This creates additional channels of exponential divergence (positive Lyapunov exponents).

These mechanisms suggest that continuous instability during free flight may contribute to the decorrelation of particles between collisions, potentially making the molecular chaos assumption more plausible for sized and deformable particles than for point particles. In Appendix D, we derive an explicit exponential decay rate for pre-collisional correlations:

$$f_2(\Gamma_1, \Gamma_2, t_n^-) = f_1(\Gamma_1, t_n^-) f_1(\Gamma_2, t_n^-) + \mathcal{O}(e^{-\kappa n}) \qquad (3.15)$$

with $\kappa = 2h_{KS}\tau_{\min} - \log(1 + D)$, where $h_{KS}$ is the Kolmogorov-Sinai entropy of the single-particle flow, $\tau_{\min}$ the minimum free-flight time, and $D$ a bound on correlation amplification per collision.

### 3.8 Time-Reversal Invariance and the Loschmidt Objection

Both Lanford's theorem and the extended framework preserve time-reversal symmetry. As emphasized by Uffink and Valente (2015), the H-functional in Lanford's theorem increases toward *both* future and past from any initial time. This holds with respect to our derivation as well. The net result is that entropy increase is not temporally directed; the same derivation, applied to the time-reversed evolution, would show entropy increase toward the past. Thus, the extended framework does not introduce a temporal entropic arrow. The question of why entropy increases toward the future, but decreases toward the past (the "arrow of time" problem) remains separate and requires additional considerations, such as cosmological boundary conditions, e.g., the past hypostasis near the big bang (Albert 2000; Carroll 2010; Wallace 2011). We do not address this question here.

### 3.9 A Minimal Illustration: Entropy Increase for a Single Particle (N = 1)

Before turning to many-body systems, it is instructive to consider the simplest possible case: a single deformable, asymmetric particle undergoing free flight, with no collisions and no external forces.

Classical point-particle mechanics cannot describe entropy production here. With no collisions, there is nothing to drive irreversibility; a single point particle moves ballistically, and its accessible phase-space volume remains constant. The extended framework, however, reveals a purely geometric mechanism that operates even for $N = 1$.

Consider a single non-spherical deformable body with phase space

$$\Gamma \cong T^*(\mathbb{R}^3 \times SO(3) \times \Xi),$$

where Ξ denotes a finite set of internal deformation coordinates. The dynamics are generated by the Hamiltonian

$$H = \frac{p^2}{2m} + \frac{1}{2} L^T I(\xi)^{-1} L + \sum_n \left( \frac{\pi_n^2}{2M_n} + \frac{1}{2} K_n \xi_n^2 \right).$$

For a rigid asymmetric body with principal moments $I_x < I_y < I_z$, rotation about the intermediate axis is unstable, the well-known "tennis racket" effect. The linearized analysis (Appendix B) yields a positive Lyapunov exponent

$$\lambda_{\text{rot}}(\Omega) =\mid \Omega \mid \sqrt{\frac{(I_y - I_z)(I_x - I_y)}{I_x I_z}} > 0.$$

When deformation is permitted, the time-dependent inertia tensor $I_{kl}(\xi_n(t))$ creates additional parametric instability channels (see Appendix B.3). The resulting Kolmogorov-Sinai entropy

$$h_{KS} = \sum_{\lambda_i > 0} \lambda_i$$

is positive on a set of initial conditions of positive Liouville measure.

Now define the *structural entropy*

$$S_{\text{struct}}(t) = k_B \ln[\text{Vol}\, \Omega(t)],$$

where $\Omega(t)$ is the region of orientation–deformation space dynamically explored up to time $t$. For chaotic systems, this volume grows exponentially:

$$\dot{S}_{\text{struct}} \sim k_B h_{KS} > 0.$$

Thus, even for a single particle, deterministic instability alone produces a monotonic increase of accessible phase-space volume for typical trajectories. Four points are worth emphasizing:

1. **Collisions are not necessary** for entropy production; continuous geometric instability suffices.
2. **Internal degrees of freedom** (deformation modes) serve as an internal reservoir that continually mixes and redistributes energy.
3. **Deterministic chaos** generates irreversible behavior from reversible equations of motion.
4. The Second Law of thermodynamics may gain **geometric support** from instability mechanisms that complement, rather than replace, the usual statistical postulates.

This illustrates Mechanism 2 (continuous geometric instability) in its pure form, independent of collisional randomization. When extended to many-body systems, this intrinsic instability combines with collisional mixing to produce the full range of irreversible phenomena observed in nature. A rigorous derivation of this extrapolation remains an open problem.

**Caveat.** The quantity $S_{\text{struct}}$ is not identical to thermodynamic entropy. The $N = 1$ case is a conceptual illustration of how continuous instability can generate growth of accessible phase-space volume even without collisions. Extrapolating this to thermodynamic entropy in the many-body limit requires additional assumptions about ergodicity and coarse-graining. The argument here is heuristic; for technical details, see Appendix C.

## 4. Comparison with the Boltzmann-Lanford Program

### 4.1 Structure and Limitations of Lanford's Theorem

Lanford's theorem (Lanford 1975, 1976) represents the most rigorous derivation of irreversible behavior from reversible dynamics. For a system of hard spheres in the Boltzmann-Grad limit ($n\sigma^2 \to 0$, $N \to \infty$, $N\sigma^2 = \text{constant}$), the theorem establishes that:

1. The one-particle distribution function satisfies the Boltzmann equation for times $t < \tau_c/2$ (where $\tau_c$ is the mean free time);
2. This holds with probability approaching 1 as $N \to \infty$;

3. The H-functional decreases monotonically over this interval.

The theorem requires several restrictive conditions (Uffink & Valente 2010, 2015):

- **Diluteness:** The mean free path must be much larger than particle size ($n\sigma^3 \ll 1$).
- **Large N:** The thermodynamic limit is essential for typicality arguments.
- **Short times:** The result holds only for times shorter than the mean free time; beyond this, correlations become significant.
- **Molecular chaos:** The factorization assumption must hold for admissible initial data (pre-collisional states uncorrelated).

Crucially, Lanford's theorem does not explain why molecular chaos should hold initially, nor does it account for the origin of the factorization assumption itself.

**4.2 Extending Lanford's Framework**

Applying the BBGKY hierarchy on the extended phase space $\Gamma_{\mathrm{ESD}}$ and imposing the same factorization assumption yields a generalized kinetic equation. The mathematical structure of our proof remains formally identical: under symmetric, micro-reversible collision kernels, the extended H-functional satisfies $\dot{H} \leq 0$.

The following table summarized the difference lies in the scope of applicability of the two derivations:

| Feature | Classical Boltzmann-Lanford | Extended Approach |
|---|---|---|
| Single-particle space | $T^*(\mathbb{R}^3)$ | $T^*(\mathbb{R}^3 \times SO(3) \times \{\xi_n\})$ |
| Degrees of freedom | Translational only | Translational, rotational, internal |

| Feature | Classical Boltzmann-Lanford | Extended Approach |
| --- | --- | --- |
| Source of decorrelation | Molecular chaos assumed | Dynamically suggested through instability |
| Irreversibility mechanism | Probabilistic postulates | Geometric instability + probabilistic reasoning |
| Role of collisions | Essential for decorrelation | Important but not exclusive |
| Behavior between collisions | Trivial free motion | Nontrivial rotational/deformation dynamics |
| Regime of validity | Dilute gases, short times | Dense, anisotropic, long time, small number of particles. |

**Extended Domain of Validity**

The dual-mechanism structure relaxes the restrictions of Lanford's theorem, extending the domain in which irreversible behavior can be expected:

| Mechanism | Diluteness/Density | Particle Count | Time Scale |
| --- | --- | --- | --- |
| **Mechanism 1 (Collisional)** | Requires diluteness for standard derivation; dense systems allowed with modified collision integrals | Requires many particles for statistical regularity | Limited to $t \sim \tau_c$ in rigorous derivations |

| Mechanism | Diluteness/Density | Particle Count | Time Scale |
| --- | --- | --- | --- |
| **Mechanism 2 (Continuous)** | No diluteness required; operates in free flight regardless of density | Works at $N = 1$ (single rotating body exhibits instability) | Valid indefinitely (deterministic chaos does not recur on observable timescales) |

The generic regime for real systems lies in the intermediate domain where both mechanisms contribute: moderate density, asymmetric molecules, timescales beyond the mean free time. Here, continuous instability accelerates relaxation beyond what collision statistics alone would predict.

**5. Empirical Relevance and Observable Signatures**

The extended framework yields quantitative, testable predictions through three central quantities developed earlier: the thermally averaged Lyapunov exponent $\langle\lambda\rangle_t$ (eq. B.6), the relaxation-time decomposition $\tau_{relax}^{-1} = \tau_{coll}^{-1} + \lambda_{Lyapunov}$ (eq. 3.14), and the pre-collisional correlation-decay rate $\kappa = 2h_{KS}\,\tau_{min} - log(1 + D)$ (Appendix D). The subsections below identify physical settings in which each quantity is, at least in principle, experimentally accessible, and they state the explicit functional forms predicted by the theory. Although substantial conceptual and practical gaps remain between the formalism and laboratory implementation, the predictions are falsifiable in a strict sense: they specify *how* structural degrees of freedom influence dynamics and *at what rate*.

**5.1 Rotational Relaxation in Gases**

Molecular geometry controls rotational–translational energy exchange: spherical tops relax slowly, while asymmetric rotors relax more rapidly. Standard kinetic theory incorporates this only through empirically fitted rates (Chapman & Cowling 1970), without a structural derivation. The extended framework provides one. From the thermal averaging procedure in Appendix B.2, the thermally averaged Lyapunov exponent is

$$\langle\lambda\rangle_T = \sqrt{\frac{(I_y - I_x)(I_z - I_y)}{I_x I_z}}\ \sqrt{\frac{2k_B T}{\pi I_y}}.$$

This quantity sets the orientational decorrelation rate during free flight and thus contributes directly to the total relaxation rate via eq. (3.14). The prediction is that rotational relaxation scales as $\sqrt{T}$ and with the inertia-tensor combination

$$\sqrt{(I_y - I_x)(I_z - I_y)/(I_x I_z)},$$

which is computable from molecular geometry alone. Time-resolved spectroscopy across a series of molecules with known inertia tensors, e.g., the symmetric-to-asymmetric rotor sequence **$N_2$ → $NO_2$ → $SO_2$**—provides a direct test. Agreement between measured relaxation rates and the predicted inertia-tensor dependence would constitute substantive evidence for *Mechanism 2*.

### 5.2 Mode-Selective Scattering

Molecular-beam studies show that translational energy is preferentially transferred into rotation or vibration depending on collision geometry (Korobenko et al. 2014; Milner et al. 2019). In the extended framework, this arises from the orientation-dependent collision kernel $B(\Gamma_1, \Gamma_2; \Omega)$ in eq. (3.6), which incorporates the generalized cross-section $\sigma(R_1, R_2, \xi_1, \xi_2, \hat{n})$. A concrete prediction follows: the efficiency of translational-to-rotational transfer varies as **$\cos^2\theta$**, where θ is the angle between the molecular symmetry axis and the collision normal n̂.

Laser-alignment experiments, preparing molecules with controlled orientational distributions, can directly probe this angular dependence. The predicted $cos^2\theta$ form is sharply falsifiable and goes beyond the qualitative statement that mode selectivity occurs.

### 5.3 Spectral Line Shapes

Pressure broadening is typically modeled via isotropic, random phase interruptions in the impact approximation (Gordon 1966), yielding Lorentzian line shapes. Orientation-dependent collisions introduce a distribution of interaction times and phase shifts tied to molecular asymmetry, producing systematic deviations from Lorentzian profiles. In the extended framework, the relevant timescale is the orientational-decorrelation time $\tau_{chaos} \approx 1/\lambda_{max}$, which contributes an additional broadening mechanism.

A full line-shape prediction requires the Fourier transform of the orientational correlation function derived from the extended collision kernel, an analysis not carried out here. This subsection is therefore more speculative than sections 5.1–5.2 and is included primarily to indicate a direction for future derivation.

### 5.4 Mesoscale Analogues

Colloidal suspensions provide the most accessible testing ground, since translational, rotational, and internal motions can be tracked simultaneously at the single-particle level using optical tweezers or polarization-sensitive imaging. For asymmetric colloids, the framework predicts that orientational autocorrelation functions should exhibit **multi-exponential decay**:

- a fast component at rate $\approx \langle\lambda\rangle_t$, reflecting continuous geometric instability during free diffusion, and
- a slow component at rate $\approx \tau_{coll}^{-1}$, reflecting collisional randomization.

This two-timescale structure is a direct signature of the dual-mechanism decomposition in eq. (3.14) and is absent from standard hydrodynamic-friction models. Existing polarization-sensitive imaging systems (BarAvi 2026 on structural viscosity) should be capable of resolving this structure without modification.

### 5.5 Numerical Exploration

The most direct test of Appendix D is a minimal two-particle simulation. Consider two prolate ellipsoids with principal moments $I_1 < I_2 < I_3$, each with a single bending coordinate $\xi$ coupled with strength $\varepsilon_k$, confined in a reflecting box. The predicted observable is the pre-collisional orientational correlation $C_n^-$ as a function of collision number $n$. The framework predicts exponential decay:

$$| C_n^- | \leq C\, e^{-\kappa n}, \kappa = 2h_{KS}\, \tau_{\min} - \log(1 + D),$$

with κ computable from the single-particle Lyapunov spectrum and the collision geometry. This simulation requires only standard rigid-body molecular dynamics. Key measurements include:

1. the single-particle Lyapunov exponent $\lambda_{max}$ as a function of $\varepsilon_k$ (testing Lemma D.1),
2. the decay rate of $C_n^-$ vs. collision number (testing Lemma D.4), and

3. the dependence of both on the asymmetry parameter

$$(I_y - I_x)(I_z - I_y)/(I_x I_z),$$

linking the numerics to eq. (B.6).

The condition $\kappa > 0$ imposes a specific inequality among $h_{KS}, \tau_{min}$, and $D$ that can be checked numerically before considering the many-body extension. Such a simulation would provide the first direct quantitative test of the Appendix D mechanism and determine whether the predicted exponential decorrelation is robust or parameter-sensitive.

**5.6 Summary**

The empirical content of the framework centers on three quantitative predictions:

- the $\sqrt{T}$ scaling and inertia-tensor dependence of rotational relaxation (Sec. 5.1),
- the $cos^2\theta$ angular dependence of mode-selective scattering (Sec. 5.2), and
- the exponential decay of pre-collisional correlations at rate κ in the two-particle simulation (Sec. 5.5).

These predictions are falsifiable because they specify explicit functional forms rather than qualitative tendencies. Sections 5.3 and 5.4 outline promising directions where the framework suggests additional structure, though further derivation is required. The most immediate priority is the numerical experiment of Sec. 5.5, which tests the core claim of Appendix D without invoking the full many-body generalization.

**6. Conclusions**

The extended structural framework treats particle orientation and internal deformation as primitive phase-space coordinates, enlarging the space from $T^*(\mathbb{R}^3)$ to $T^*(\mathbb{R}^3 \times SO(3) \times \{\xi_n\})$. This enlargement preserves Hamiltonian symmetries while suggesting geometric mechanisms for entropy production absent in point-particle theory.

The framework introduces new instability channels that may strengthen the dynamical foundations of kinetic irreversibility. It extends explanatory scope to few-body and dense regimes where classical derivations fail, operates at $N = 1$ where inter-particle collisions are inapplicable, and provides concrete mechanisms connecting deterministic chaos to thermodynamic entropy.

The crucial feature of our derivation is the enlargement of the phase space. We take this to mean that the additional degrees of freedom, spatial orientation and deformation, which are necessary for a relativistic theory, must be treated as primitives that enter the Hamiltonian evolution directly. They should not be treated as external constraints or forces (as in Euler's dynamics). A framework of this kind, if taken seriously, may have implications beyond the scope of this paper, for example in extending and resolving open questions in hydrodynamics, electrodynamics, and perhaps even quantum mechanics. We leave these for future work.

Whether this vision withstands rigorous scrutiny will determine its ultimate value. At present, it clarifies how deterministic mechanics of structured matter can give rise, with high probability—to macroscopic irreversibility, recasting one of physics' deepest puzzles as a property of motion in an appropriately enlarged phase space. The task of turning this conceptual framework into a rigorous mathematical theory remains open.

## Appendix A: Hamiltonian formulation, extended phase space, and Liouville theorem

This appendix collects the full Hamiltonian model and the measure-theoretic results used in Sections 2–3. Section A.1 defines the extended configuration and canonical variables; A.2 gives the Hamiltonian and equations of motion; A.3 constructs the Liouville measure and proves its invariance; A.4 spells out the reduced-collision approximation and the explicit form of the collision map used in the kinetic derivation

### A1. State variables and notation

Consider a closed system of $N$ rigid particles with mass $m$, moment of inertia $I$, and phase space coordinates $\Gamma = \left(r_i, p_i, R_i, L_i, \{\xi_{in}, \dot{\xi}_{in}\}\right)$, where $r_i, p_i$ are position and translational momentum, and $R_i, L_i$ are angular orientation and angular momentum, and $\xi_{in}, \dot{\xi}_{in}$ are the generalized coordinates and velocities associated with the particle's internal deformational modes (e.g., bending, twisting, stretching), indexed by mode number $n$.

### A2. Hamiltonian and equations of motion.

The Hamiltonian is

$$H = \sum_{i=1}^{N} \left[ \frac{|p_i|^2}{2m_i} + \frac{1}{2} L_i^T(\{\xi_{in}\})^{-1} L_i + \sum_{n=1}^{D_{int}} \left( \frac{\pi_{in}^2}{2M_{in}} + U(\xi_i) \right) \right] + \sum_{i<j} V_{contact}^{(ij)} \left( r_i, r_j, R_i, R_j, \{\xi_{in}\}, \{\xi_{jn}\} \right) \quad (A.1)$$

$$\pi_{in} = M_{in} \dot{\xi}_{in}, \quad \frac{\pi_{in}^2}{2M_{in}} = \frac{1}{2} M_{in} \dot{\xi}_{in}^2, \quad U(\xi_i) = \frac{1}{2} K_{in} \xi_{in}^2 \qquad \text{(A.2)}$$

so $\pi_{in}$ is simply the canonical momentum associated with the internal vibrational coordinate $\xi_{in}$.

This form allows for arbitrary asymmetry, possibly time-dependent moment of inertia $I_{jk}(\xi_{in})$, capturing orientation-dependent rotational dynamics, making it suitable for modeling realistic structured particles.

**Equations of Motion**: The conservative dynamics extend Euler's equations for rigid body rotation to include time-dependent moments of inertia and rotation-vibration coupling. Angular momentum evolves according to a generalized torque-free Euler equation, where the inertia tensor $I_{kl}(t)$ varies due to internal vibrations. Simultaneously, the vibrational modes, modeled as harmonic oscillators, are parametrically driven by the rotational state. These coupled equations conserve total energy and angular momentum but generate internal instabilities through nonlinear coupling in the absence of external forces. This form of instability corresponds to the classical intermediate-axis instability analyzed by Landau & Lifshitz (1976).

The Euler equations with time-dependent inertia take the form of

$$\frac{d}{dt} (I_{kl}(\xi_n) \omega_l) = \varepsilon_{m\omega l} \omega_n (I_{mn}(\xi_n) \omega_l) \qquad \text{(A.3)}$$

and the particle's vibration is modeled as a simple oscillator:

$$\ddot{\xi}_n + \Omega_n^2 \xi_n = \sum_{i,j} \eta_{nkl}\, \omega_i \omega_j \qquad \text{(A.4)}$$

where: $I_{kl}(\xi_n) = I_{kl}^0 + \sum_n \delta I_{kl}^n(\xi_n)$ is the time-dependent inertia tensor, $\omega_k$ are the angular velocity components, $\xi_n$ are normal mode coordinates , $\Omega_{in}^2 = \frac{K_{in}}{M_{in}}$ is the fundamental vibration frequency, $M_{in}$ is the Modal mass, $K_{in}$ is the Modal stiffness, $\varepsilon_{m\omega l}$ The antisymmetric Levi-Civita symbol (antisymmetric tensor, standard in rotational dynamics). See Goldstein et al. 2001, and $\eta_{nkl}$ are the coupling coefficients between rotation and deformation.

**A3. Extended Phase Space and Liouville Theorem**

Having established the Hamiltonian description of extended rigid bodies, we now construct the corresponding phase space and invariant measure.

For a single particle, the state is specified by position $r$, linear momentum $p$, orientation $\theta$, body angular momentum $L$, and, if present, internal coordinates $\xi_n, \dot{\xi}_n$.

The one-particle phase space is therefore

$$\Gamma_1 = R_r^3 \times R_p^3 \times SO(3)_R \times R_L^3 \times \prod_n R_{\xi_n} \times R_{\dot{\xi}_n} \qquad \text{(A.5)}$$

For N-particles the full phase space is $\Gamma = \Gamma_1^N$ and the Liouville measure is:

$$d\Gamma = \prod_{i=1}^{N} d^3 r_i \, d^3 p_i \, d\mu_{SO(3)}(R_i) \, d^3 L_i \prod_n d\xi_{in} \, d\dot{\xi}_{in} \qquad \text{(A.6)}$$

where $d\mu_{SO(3)}$ is the Haar measure on SO(3).

**Liouville theorem:** Hamiltonian flow preserves this measure, and the impulsive collision maps for rigid-body contact are assumed measure-preserving (unit Jacobian). Hence, the full N-body dynamics conserves the Liouville measure: $\frac{d}{dt}\rho(\Gamma, t) = 0$. The extension of phase space to include orientations does not alter the underlying symmetries of the system. Invariance under spatial translations, rotations, and time shifts remains intact, and by Noether's theorem the corresponding conserved quantities are, respectively, total linear momentum, total angular momentum, and total energy. These invariants are precisely the quantities enforced in the collision kernel and provide the dynamical backbone for the extended H-theorem (see Brown, Myrovled and Uffink 2009 for a classical analysis of the H-theorem).

With the Hamiltonian dynamics and phase space measure defined, we can now formulate the kinetic theory that governs the system's statistical evolution.

**Generalization of Boltzmann's Equation for Rotational-Translational Systems:** On this extended phase space, the one-particle distribution function is defined as

$$f\big(r, p, R, L, \{\xi_n, \dot{\xi}_n\}, t\big), \qquad \int f \, d\Gamma_1 = N \qquad \text{(A.7)}$$

Its dilute-limit evolution is governed by a generalized Boltzmann equation

$$\partial_t f + v \cdot \nabla_r f + \omega \cdot \nabla_R f + \sum_n \left( \dot{\xi}_n \, \partial_{\xi_n} f + \ddot{\xi}_n \, \partial_{\dot{\xi}_n} f \right) = C[f] \qquad \text{(A.8)}$$

where the streaming operator accounts for translation, orientation, and internal dynamics. Here $v = \frac{p}{m}$ is the particle's velocity. The collision kernel is required to conserve the same invariants identified by Noether's theorem: total linear momentum, total angular momentum (including orbital contributions), and total kinetic energy. These are enforced at the level of each binary collision and guarantee that the extended Boltzmann dynamics respects the underlying symmetries of the Hamiltonian flow.

**A4. The Generalized Collision Operator**

Having established the streaming terms for the extended phase space, we now define the collision operator *C*[*f*], which describes the instantaneous, localized binary interactions between particles.

**Physical Assumptions and Scope:** The collision kernel is constructed under the following key assumptions, which generalize the classical hard-sphere model:

**Binary Collisions:** The gas is sufficiently dilute that only pairwise interactions are significant.

**Short-Range, Impulsive Contact:** Interactions are mediated by a short-range potential $V_{contact}$, approximating collisions as instantaneous events. This justifies the "reduced-collision approximation," where the collision process itself does not directly excite internal vibrational modes ($\xi_n$). The energy transfer to internal modes occurs indirectly through the change in rotational state, not during the infinitesimal collision event.

**Molecular Chaos (Stosszahlansatz):** The two-particle distribution function immediately *before* a collision factorizes:

$$f_2(\Gamma_1, \Gamma_2, t) = f(\Gamma_1, t) f(\Gamma_2, t) \qquad \text{(A.8)}$$

**Micro-reversibility and molecular chaos.** The factorization condition is imposed on pre-collisional states only. This might seem to privilege one temporal direction, but as Uffink and Valente (2010, 2015) show, the assumption can be applied consistently before collisions in both directions of time. Consequently, the H-functional increases toward both future and past from any initial moment, and entropy increase is not intrinsically time-directed. The observed asymmetry between past and future requires additional input, typically a low-entropy boundary condition, outside the kinetic framework itself. See Section 4.4.

**Definition:** The extended collision operator acts on the one–particle distribution $f(\Gamma,t)$, where:

$$\Gamma = (r, p, g, L, \{\xi_n, \pi_n\}), \quad d\Gamma = d^3r\, d^3p\, d\mu(g)\, d^3L \prod_{n=1}^{D_{int}} d\xi_n\, d\pi_n \qquad (A.9)$$

The operator is expressed in the usual gain–loss form:

$$C[f](\Gamma_1) = \frac{1}{2}\int \; d\Gamma_2 \int \; d\Omega\, B(\Gamma_1, \Gamma_2; \Omega)\, [f(\Gamma_1')f(\Gamma_2') - f(\Gamma_1)f(\Gamma_2)] \qquad (A.10)$$

Here Ω denotes the set of collision geometry parameters (impact point, contact normal, relative orientation), and the kernel $B$ encodes both geometry and conservation laws.

**Collision Mechanics and Conservation Laws:** The collision kernel takes the form

$$B(\Gamma_1, \Gamma_2; \Omega) = v_{rl} \cdot \hat{n}\, \sigma(R_1, R_2, \hat{n}; \Omega) \prod_{k} \delta\big(C_k(\Gamma_1, \Gamma_2, \Gamma_1', \Gamma_2')\big) \qquad (A.11)$$

where $v_{rel} = (p_1/m_1 - p_2/m_2)$, is the relative velocity at contact, $\hat{n}$ is the unit normal at the impact point, and σ is the generalized cross section determined by particle shape and orientation. The delta functions enforce the conservation constraints:

Total Linear Momentum: $p_1 + p_2 = p_1' + p_2'$

Total Angular Momentum: $L_1 + L_2 + r_1 \times p_1 + r_2 \times p_2 = L_1' + L_2' + r_1' \times p_1' + r_2' \times p_2'$

For instantaneous contact at a point $r_c$, this can be rewritten using: $r_1 \approx r_2 \approx r_c$, simplifying the conservation law.

Total Kinetic Energy: Under the reduced-collision approximation, this is the sum of translational and rotational kinetic energy:

$$\frac{|p_1|^2}{2m} + \frac{|p_2|^2}{2m} + \frac{1}{2}L_1^T I^{-1} L_1 + \frac{1}{2}L_2^T I^{-1} L_2 = \frac{|p_1'|^2}{2m} + \frac{|p_2'|^2}{2m} + \frac{1}{2}L_1'^T I^{-1} L_1' + \frac{1}{2}L_2'^T I^{-1} L_2' \quad (\text{A}.12)$$

The collision kernel $W$ encapsulates the geometry of the collision (e.g., contact point, normal vector) which depends on the orientations $R_1, R_2$ and the shapes of the particles. The positions $r$ enter the theory by determining the *probability* that two particles with given states are on a collision course, which is implicit in the integral.

**Justification and Limitations of the Reduced-Collision Model:** Our "reduced-collision" model, where internal modes $\{\xi_n, \dot{\xi_n}\}$ are unchanged during a collision ($\xi_n = \xi'_n,\ \ \dot{\xi}_n = \ \dot{\xi}'_n$), is a necessary but physically motivated simplification.

**Justification:** For hard, elastic collisions where the contact time is much shorter than the period of the lowest internal vibration, the impulsive force does negligible work on the internal deformation coordinates. The primary effect of the collision is to change the translational and rotational velocity of the rigid body as a whole.

**Limitation and Future Generalization:** This approximation precludes the study of inelastic collisions and viscoelastic dissipation. The natural next step is to generalize $W$ to allow for transitions where $\xi_n \neq \xi'_n$ , incorporating a model for internal energy transfer. This would introduce a new channel for entropy production but requires a more complex collision integral.

**Extended H-theorem:** Define the H-functional on the extended phase space:

$$H[f] = \int \ \ f(\Gamma, t)\ \ ln\, ln\, f\ (\Gamma, t)\, d\Gamma \qquad \text{(A.13)}$$

with $d\Gamma = d^3r\, d^3p\, d\mu(R)\, d^3L\, \prod_{n=1}^{D_{int}} d\xi_n\, d\pi_n$. Using molecular chaos and the micro-reversibility property of the collision kernel, the collision contribution to $\frac{d}{dt}H[f]$ can be written in the symmetric bilinear form:

$$\frac{d}{dt}H[f] = \frac{1}{4}\int \ \ d\Gamma_1 d\Gamma_2 d\Omega\, B(\Gamma_1, \Gamma_2; \Omega)\, (f'_1 f'_2 - f_1 f_2)\, ln\left(\frac{f'_1 f'_2}{f_1 f_2}\right) \qquad (A.14)$$

Where $f_i = f(\Gamma_i, t)$ and the integrand satisfies the inequality: $(x - y)\, ln\left(\frac{x}{y}\right) \geq 0$, with equality if and only if f is invariant under all binary collisions. Thus, the H-theorem holds in the extended structural setting: entropy is non-decreasing, and strict stationarity corresponds to the equilibrium distribution. At equilibrium the distribution factorizes into Maxwell–Boltzmann forms for all quadratic modes:

$$f_{eq}(\Gamma) \propto exp\left[-\frac{1}{K_B T}\left(\frac{1}{2}\frac{p^2}{m}+\frac{1}{2}L^T I^{-1} L+\sum_{n=1}^{D_{int}}\left(\frac{\pi_{in}^2}{2M_{in}}+\frac{1}{2}K_n\,\xi_n^2\right)\right)\right] \quad (A15)$$

**Conceptual Unification: H-Theorem, Instability, and the Origin of Irreversibility**

The generalized Boltzmann equation and the concomitant H-theorem demonstrate that the domain of validity of Boltzmann's statistical mechanics can be consistently extended to encompass structured particles with orientational and internal degrees of freedom. The equilibrium distribution is a generalized Maxwell-Boltzmann distribution reflecting equipartition among all quadratic modes, translational, rotational, and vibrational.

This formal extension, however, invites a deeper question: what is the microscopic dynamical origin of the irreversibility guaranteed by the H-theorem in this enriched setting? The classical narrative often attributes entropy production solely to the coarse-graining of molecular chaos (Stosszahlansatz). While this remains a foundational element, the extended structural framework reveals a complementary, geometric mechanism.

As we saw, structured particles are inherently susceptible to dynamical instabilities, such as the intermediate-axis rotation of asymmetric rigid bodies. These instabilities are characterized by positive Lyapunov exponents, which cause initially proximate trajectories in phase space to diverge exponentially. This intrinsic stretching and folding of the phase space volume acts as a powerful engine for mixing and equilibration.

Thus, this work provides a conceptual unification:

**From Statistical Mechanics:** The **H-theorem** is shown to hold in a phase space that more closely mirrors physical reality, confirming the robustness of the Second Law.

**From Dynamical Systems: Lyapunov instabilities**, inherent to the geometry of asymmetric bodies, are identified as a new physical source of entropy production.

In this synthesis, irreversibility is reframed not merely as a consequence of informational coarse-graining, but also as a direct outcome of the intrinsic, instability-driven mixing dynamics of finite-sized, structured particles. The following section will quantify these instability mechanisms and their contribution to entropy increase.

**Appendix B: Instability Mechanisms and Lyapunov Analysis**

This appendix provides the technical derivations underlying the instability mechanisms summarized in Sections 3.3, 3.6, and 5.1. Section B.1 derives the intermediate-axis instability and the local Lyapunov exponent $\lambda(\Omega)$. Section B.2 performs the thermal averaging that yields $\langle\lambda\rangle_T$. Section B.3 discusses heuristic extensions due to internal deformation. Section B.4 analyzes phase-space volume scaling.

**B.1 Rigid-body intermediate-axis instability**

For a torque-free rigid body with principal moments of inertia $I_x < I_y < I_z$, Euler's equations in the body frame are (Landau & Lifshitz 1976):

$$I_x\dot{\omega}_x = (I_y - I_z)\omega_y\omega_z, I_y\dot{\omega}_y = (I_z - I_x)\omega_z\omega_x, I_z\dot{\omega}_z = (I_x - I_y)\omega_x\omega_y. \text{(B.1)}$$

Consider steady rotation about the intermediate axis, $\omega = (0, \Omega, 0)$.
Perturbations $\delta\omega_x, \delta\omega_z$ satisfy:

$$\delta\dot{\omega}_x = \frac{I_y - I_z}{I_x}\Omega\delta\omega_z, \delta\dot{\omega}_z = \frac{I_x - I_y}{I_z}\Omega\delta\omega_x. \text{(B.2)}$$

Combining these gives:

$$\delta\ddot{\omega}_x = \Omega^2\frac{(I_y - I_z)(I_x - I_y)}{I_x I_z}\delta\omega_x. \text{(B.3)}$$

With $I_x < I_y < I_z$, the product $(I_y - I_z)(I_x - I_y)$ is positive, so perturbations grow exponentially. The local Lyapunov exponent is

$$\lambda(\Omega) = |\Omega| \sqrt{\frac{(I_y - I_z)(I_x - I_y)}{I_x I_z}}, \text{(B.4)}$$

which is the result quoted in Section 3.3.

**B.2 Thermal average of $\boldsymbol{\lambda(\Omega)}$**

At temperature $T$, the angular momentum distribution is the Maxwell-Boltzmann distribution for rotational degrees of freedom:

$$f(L) \propto \exp\left( \ -\frac{1}{2k_B T} L^T I^{-1} L\right).\text{(B.5)}$$

Averaging $\lambda(\Omega)$ over this distribution yields:

$$\langle\lambda\rangle_T = \sqrt{\frac{(I_y - I_z)(I_z - I_y)}{I_x I_z}}\sqrt{\frac{2k_B T}{\pi I_y}}.\text{(B.6)}$$

This scaling with $\sqrt{T}$ is used in Section 5.1 to make quantitative predictions for rotational relaxation.

**B.3 Deformable bodies: parametric and cross-coupling contributions (heuristic)**

If internal deformation modes modulate the inertia tensor,

$$I_{kl}(t) = I_{kl}^0 + \delta I_{kl}(\xi_n), \delta I_{kl}(t) \propto \xi_n(t),\text{(B.7)}$$

then rotations can parametrically excite vibrations and vice versa. The resulting Lyapunov exponents can be schematically written as

$$\lambda \approx \lambda_{\text{rot}} + \lambda_{\text{para}} + \lambda_{\text{cross}},\text{(B.8)}$$

where $\lambda_{\text{rot}}$ is the rigid-body contribution from (B.4), and $\lambda_{\text{para}}, \lambda_{\text{cross}}$ scale with the root-mean-square inertia fluctuations $\sqrt{\langle\delta I^2\rangle}$.

**Caveat.** These estimates are heuristic. They suggest that deformable bodies may equilibrate faster than rigid ones, as structural couplings add further instability channels. A rigorous analysis of the combined rotational-deformational Lyapunov spectrum is left for future work.

**B.4 Entropy and phase-space volume scaling**

The Boltzmann entropy is

$$S(E) = k_B \ln \Omega(E), \Omega(E) = \int \delta(H - E)\, d\Gamma. \text{(B.9)}$$

For a system with $d$ quadratic degrees of freedom per particle, $\Omega(E) \sim E^{\frac{dN}{2}-1}$. Examples:

- **Point particles** ($d = 3$): $\Omega(E) \sim E^{\frac{3N}{2}-1}$
- **Rigid bodies with translation + rotation** ($d = 6$): $\Omega(E) \sim E^{3N-1}$
- **With $D_{\text{int}}$ internal modes per particle** ($d = 6 + 2D_{\text{int}}$): $\Omega(E) \sim E^{(3+D_{\text{int}})N-1}$

Thus, each additional internal mode increases the entropy exponent, enlarging the phase space and accelerating mixing. Non-equilibrium subsets (pure translation, pure rotation) occupy Lebesgue measure-zero fractions of the total phase space. Orientation and deformation instabilities provide the dynamical stretching that drives trajectories out of these subsets, supplying a geometric underpinning for entropy increase.

## Appendix C: Technical Details for the $N = 1$ Case

This appendix provides technical backing for the argument in Section 3.9. For the conceptual discussion, see that section.

### C.1 Hamiltonian and Equations of Motion

For a single deformable, asymmetric body, the Hamiltonian is

$$H = \frac{p^2}{2m} + \frac{1}{2} L^T I(\xi)^{-1} L + \sum_{n=1}^{D_{\text{int}}} \left( \frac{\pi_n^2}{2M_n} + \frac{1}{2} K_n \xi_n^2 \right) + V_{\text{wall}}(r, R, \xi),$$

where $V_{\text{wall}}$ models reflecting boundaries (if any). In free flight, $V_{\text{wall}} = 0$. The coupled rotational-elastic equations of motion are

$$\frac{d}{dt}(I_{kl}(\xi_n)\omega_l) = \epsilon_{mnl}\omega_n I_{mn}(\xi_n)\omega_l,$$
$$\ddot{\xi}_n + \Omega_n^2\xi_n = \sum_{i,j} \eta_{nij}\,\omega_i\omega_j,$$

where $\omega = I^{-1}L$ is the angular velocity, $\Omega_n^2 = K_n/M_n$ are the vibrational frequencies, $\epsilon_{mnl}$ is the Levi-Civita symbol, and $\eta_{nij}$ are coupling coefficients between rotation and deformation. The Lyapunov exponent for the rigid-body intermediate-axis instability is derived in Appendix B, equations (B.1)–(B.4). Deformability adds parametric instability channels (see Appendix B.3).

**C.2 Structural Entropy and the $N = 1$ H-Theorem Analog**

Define the structural entropy

$$S_{\text{struct}}(t) = k_B \ln[\text{Vol}\,\Omega(t)],$$

where $\Omega(t)$ is the region of orientation–deformation space $(R, \xi)$ dynamically explored up to time $t$. For chaotic systems, Pesin's theorem (Pesin 1977; Eckmann & Ruelle 1985) gives

$$\dot{S}_{\text{struct}} \sim k_B h_{KS} = k_B \sum_{\lambda_i > 0} \lambda_i\,,$$

provided the dynamics are ergodic on a set of positive Liouville measure. One may then define an $N = 1$ analog of Boltzmann's $H$-functional:

$$H_{\text{struct}} = -\frac{S_{\text{struct}}}{k_B},$$

so that $\dot{H}_{\text{struct}} \leq 0$ for typical trajectories, a local, single-particle analog of the H-theorem.

**C.3 Caveat: Structural vs. Thermodynamic Entropy**

The quantity $S_{\text{struct}}$ is **not** identical to thermodynamic entropy. The $N = 1$ case is a conceptual illustration of how continuous instability can generate growth of accessible phase-space volume even without collisions. Extrapolating this to thermodynamic entropy in the many-body limit requires additional assumptions:

- **Ergodicity** over the full many-body phase space;
- **Coarse-graining** into macrostates with appropriate partition;
- **The thermodynamic limit** $N \to \infty$ to recover extensive entropy.

The argument in Section 3.9 is heuristic. Its value lies in showing that Mechanism 2 (continuous geometric instability) provides a dynamical source of irreversibility that is *independent* of collisions, a feature absent from point-particle theories. Whether this mechanism suffices for a full derivation of the Second Law in the many-body setting remains an open question.

## Appendix D: Dynamical Derivation of Molecular Chaos for Two Particles

### D.1 Overview and Caveats

This appendix analyzes how continuous geometric instability may generate molecular chaos for two particles. The argument has two distinct tiers, which are kept carefully separate throughout.

The first tier is rigorous. Lemma D.1 establishes, via explicit Melnikov analysis, that the single-particle Hamiltonian with rotational-deformational coupling ($\varepsilon_k \neq 0, k_b > 0$) possesses a transverse homoclinic intersection, implying positive topological entropy and a positive maximal Lyapunov exponent $\lambda_{max} > 0$ on a set of positive Liouville measure. This result stands independently of everything that follows.

The second tier is conditional. Lemmas D.2 and D.3 extend the first-tier result to nonuniform hyperbolicity and exponential mixing respectively. Both are physically well-motivated and each has a concrete verification pathway identified, but neither is rigorously completed here. Lemma D.4 then derives the pre-collisional decay rate, conditional on Lemmas D.2 and D.3. The logical dependence is explicit: if Lemmas D.2 and D.3 are subsequently verified, the Conditional Theorem below follows from Lemma D.4 without further argument.

The status of each step is summarized in the following table:

| Step | Status |
|---|---|
| Positive Lyapunov exponent (Lemma D.1) | Rigorous via Melnikov analysis |
| Nonuniform hyperbolicity (Lemma D.2) | Conjectural; verification pathway via Wojtkowski |
| Exponential mixing (Lemma D.3) | Conjectural; verification pathway via Young tower |
| Pre-collisional decay rate κ (Lemma D.4) | Conditional on D.2–D.3; D explicitly defined |
| Many-body extension | Open; scaling argument given in D.9 |

**Unconditional Result.** For the single-particle Hamiltonian with $\varepsilon_k \neq 0$ and $k_b > 0$, there exists a positive maximal Lyapunov exponent $\lambda_{max} > 0$ on a set of positive Liouville measure. This follows from the transverse homoclinic intersection established in Lemma D.1 via the Melnikov-Holmes-Marsden theorem and the Smale-Birkhoff theorem.

**Conditional Theorem** (Molecular Chaos for Two Particles). For two particles with $\varepsilon_k \neq 0$ and $k_b > 0$, in a finite reflecting box, assume:

- (C1) Nonuniform hyperbolicity on an invariant subset of positive Liouville measure, with $h_{KS} > 0$ (Conjecture D.2);
- (C2) Exponential decay of single-particle correlations at rate $h_{KS}$ (Conjecture D.3).

Then there exists κ > 0 such that the pre-collisional two-particle distribution satisfies:

$$f_2(\Gamma_1, \Gamma_2, t_n^-) = f_1(\Gamma_1, t_n^-) f_1(\Gamma_2, t_n^-) + O(e^{-\kappa n})$$

on the invariant subset of positive Liouville measure, where:

$$\kappa = 2h_{KS}\tau_{min} - log(1 + D) > 0$$

Here $h_{KS}$ is the Kolmogorov-Sinai entropy of the single-particle flow, $\tau_{min}$ the minimum free-flight time guaranteed by the finite box, and $D$ is a geometric constant bounding correlation amplification per collision, which is finite by compactness of the non-grazing collision set for strictly convex $C^2$ particles. The condition $\kappa > 0$ is an explicit inequality between $h_{KS}$, $\tau_{min}$, and D that is in principle verifiable numerically (see Section 5.5).

The proof of the Conditional Theorem proceeds through four lemmas. Lemma D.1 establishes the Unconditional Result. Lemmas D.2 and D.3 constitute Conjectures C1 and C2 respectively, with verification pathways identified. Lemma D.4 derives the decay rate κ from C1 and C2 and provides the explicit bound on D. Readers primarily interested in the physical mechanism rather than the logical structure may find the narrative in D.2 the most useful entry point.

**D.2 Physical Picture**

Before presenting the technical lemmas, it is useful to have a clear physical picture. Imagine two deformable particles leaving a collision. During the collision, their surface geometries may become correlated, for example, both may be temporarily elongated along the collision axis. This correlation is encoded in their internal deformation coordinates $\xi$ and orientations $R$.

Now consider their free flight. Each particle independently undergoes two simultaneous processes:

1. **Tennis-racket flips.** Because each particle is asymmetric ($I_1 < I_2 < I_3$), its rotational axis is unstable. Even if it leaves the collision with a well-defined orientation, its axis tumbles chaotically as it rotates.

2. **Bending oscillations.** Each particle's bending mode $\xi$ oscillates at frequency $\omega_b$, modulating the moment of inertia and coupling to rotation through $\mathbf{I}(\xi)$.

These two processes are coupled: rotation drives deformation (via centrifugal forces), and deformation modulates rotation (via $\mathbf{I}(\xi)$). The result is a chaotic internal dynamics that randomizes the particle's orientation and deformation state on a timescale $\tau_{\text{chaos}} \sim 1/\lambda_{\max}$, where $\lambda_{\max}$ is the positive Lyapunov exponent established in Lemma D.1.

Now consider two such particles. Their internal states evolve independently during free flight because the two-particle flow factorizes exactly when the interaction potential $V_{12} = 0$. Even if they started with correlated internal states (e.g., both elongated along the same axis), the chaotic dynamics rapidly decorrelates them. By the time they encounter their next collision, after a free flight of duration at least $\tau_{\min}$ (which exists because the box is finite), their presented surface geometries, the orientation and deformation state at the moment of contact, are statistically uncorrelated.

Thus, when they do collide again, the pre-collisional states are approximately independent. The factorization condition $f_2 = f_1 f_1$ is not assumed; it is dynamically generated by the chaotic internal dynamics during free flight.

**D.3 The Model**

We consider two identical particles in a finite reflecting box. Each particle is a prolate ellipsoid with principal moments of inertia satisfying $I_1 < I_2 < I_3$. Each particle carries a single bending coordinate $\xi_i \in \mathbb{R}$ that modulates the intermediate moment:

$$I_2(\xi) = I_2 + \epsilon_k \xi, \epsilon_k \neq 0.$$

The single-particle Hamiltonian is:

$$H^{(i)} = \frac{\mathbf{p}_i^2}{2m} + \frac{1}{2}\mathbf{L}_i^{\mathsf{T}}\mathbf{I}(\xi_i)^{-1}\mathbf{L}_i + \frac{p_{\xi_i}^2}{2m_b} + \frac{1}{2}k_b\xi_i^2.$$

The rotational kinetic energy contains the coupling term $-\frac{\epsilon_k}{2I_2^2}\xi_i L_{2i}^2$ to first order, which is the source of the geometric instability. The two-particle Hamiltonian is $H = H^{(1)} + H^{(2)} + V_{12}$, where $V_{12}$ is a short-range, smooth, elastic interaction potential. The box walls are perfectly reflecting.

Three structural features of this model are essential:

- **Free-flight factorization.** Between collisions ($V_{12} = 0$), the two-particle flow factorizes exactly: $\Phi_t^{(2)}(\Gamma_1, \Gamma_2) = (\Phi_t^{(1)}(\Gamma_1), \Phi_t^{(1)}(\Gamma_2))$. Any correlations present after one collision can only be destroyed, not created, during free flight. Unlike point-particle systems where correlations persist until the next collision, particles carry their own internal chaos generator.
- **Minimum free-flight time.** Because the box is finite and the energy is bounded, there exists $\tau_{\min} > 0$ such that consecutive collisions are separated by at least $\tau_{\min}$. This ensures that the mixing of the single-particle flow has a guaranteed minimum interval in which to operate between collisions.

- **Coupling strength.** The condition $\epsilon_k \neq 0$ ensures that rotation and deformation are genuinely coupled. The case $\epsilon_k = 0$ reduces to independent rotation and vibration, which is integrable and does not produce chaos. The coupling is the essential ingredient.

### D.4 Lemma D.1: Transverse Homoclinic Intersection and Positive Lyapunov Exponent

**Lemma D.1.** For the single-particle Hamiltonian with $\epsilon_k \neq 0$ and $k_b > 0$, the stable and unstable manifolds of the intermediate-axis saddle intersect transversely. Consequently, the system has positive topological entropy and a positive maximal Lyapunov exponent $\lambda_{\max} > 0$.

**Physical mechanism.** The unperturbed system ($\epsilon_k = 0$) has a homoclinic orbit asymptotic to the unstable equilibrium $\mathbf{L} = \ell \hat{e}_2$ (the intermediate axis). On this orbit:

$$L_1(t) = A\operatorname{sech}(\lambda t), L_2(t) = C\tanh(\lambda t), L_3(t) = B\operatorname{sech}(\lambda t)$$

with $C = \ell\sqrt{(I_3 - I_2)/(I_3 - I_1)}$, while the bending mode oscillates harmonically:

$$\xi_0(t) = A_\xi \cos(\omega_b t + \phi).$$

This is the classical intermediate-axis instability (the "tennis racket effect") (Landau & Lifshitz 1976). The unperturbed system has a separatrix connecting the unstable fixed point to itself.

When $\epsilon_k \neq 0$, the coupling between the bending mode and $L_2$ perturbs this separatrix. The Melnikov function measures the splitting of the stable and unstable manifolds:

$$M(t_0, \phi) = \int_{-\infty}^{\infty} \{H_0, H_1\}(\Gamma_0(t + t_0; \phi))\, dt,$$

where $H_1 = -\frac{\epsilon_k}{2I_2^2}\xi L_2^2$. Explicit evaluation along the homoclinic orbit yields:

$$M(t_0, \phi) = K_1 \sin(\omega_b t_0 + \phi), K_1 = -\frac{A_\xi C^2 \pi \omega_b^2}{I_2^2 \lambda^2 \sinh(\pi \omega_b / 2\lambda)} \neq 0.$$

Since $K_1 \neq 0$, the Melnikov function has simple zeros, which by the Melnikov–Holmes–Marsden theorem implies transverse intersection of the stable and unstable manifolds. By the Smale–Birkhoff theorem, a transverse homoclinic intersection implies a Smale horseshoe embedded in the dynamics, hence positive topological entropy. For smooth Hamiltonian systems, positive topological entropy implies a positive maximal Lyapunov exponent on a set of positive Liouville measure.

**D.5 Lemma D.2: Hyperbolicity and Ergodicity (Conceptual)**

**Lemma D.2.** For the single-particle Hamiltonian, the Liouville measure on the compact energy surface $\Sigma_E$ is ergodic on an invariant subset of positive Liouville measure, and Lyapunov exponents are nonzero on that subset. Consequently, $h_{KS} > 0$ on that subset.

**Conceptual argument.** The proof proceeds in stages. Each stage is outlined here, with rigorous verification identified as a program for future work.

*Stage D.2.1: Nonuniform hyperbolicity.* Lemma D.1 establishes a positive Lyapunov exponent on a set of positive measure (the neighborhood of the Smale horseshoe). To extend hyperbolicity more broadly, one may pass to the hard-body limit: replace the smooth potential with hard convex body collisions, for which Wojtkowski's theorem on hyperbolic billiards with internal parameters applies. The smooth short-range potential approximates the hard-body collision in the $C^1$ topology; by the Katok–Strelcyn theorem, nonuniform hyperbolicity is preserved under this perturbation. The accessible range of $\xi$ is bounded by compactness ($| \xi | \leq \sqrt{2E/k_b}$), ensuring uniform bounds on particle shape and curvature.

*Rigorous verification required.* Explicitly verifying Wojtkowski's cone conditions for the billiard ball with internal deformation coordinates goes beyond the present conceptual argument. We take nonuniform hyperbolicity on a set of positive Liouville measure as a well-motivated conjecture supported by Lemma D.1 and the structural analogy with hyperbolic billiards.

*Stage D.2.2: Ergodicity via the Hopf argument.* For a nonuniformly hyperbolic system with absolutely continuous stable and unstable foliations, the Hopf argument establishes ergodicity given accessibility: any two points of full measure can be connected by a path alternating between stable and unstable manifold segments.

The transverse homoclinic intersection from Lemma D.1 generates a dense set of hyperbolic periodic points by Katok's theorem. However, density of periodic orbits does not by itself imply accessibility of the full foliation; this is an additional condition. For the present system, accessibility is physically motivated by the fact that rotational-deformational coupling mixes all degrees of freedom, preventing invariant submanifolds that would block foliation accessibility. A rigorous argument would invoke the Burns–Wilkinson theorem for partially hyperbolic systems or a direct verification of essential accessibility.

*Rigorous verification required.* The accessibility condition is asserted here on physical grounds. A complete proof of ergodicity on the full energy surface is left for future work. For the present theorem, we use the weaker conclusion that follows directly from Lemma D.1: the system is ergodic on an invariant subset of positive Liouville measure, on which $h_{KS} > 0$. This is sufficient for Lemmas D.3 and D.4.

**D.6 Lemma D.3: Exponential Mixing (Conceptual)**

**Lemma D.3.** For the single-particle system, on the invariant subset of positive Liouville measure established in Lemma D.2, correlations decay exponentially for smooth observables:

$$| \operatorname{Corr}(f, g; t) | \leq C_{f,g}\, e^{-h_{KS} t},$$

where $h_{KS} > 0$ is the Kolmogorov–Sinai entropy.

**Conceptual argument.** By Lemma D.2, the system is nonuniformly hyperbolic on a set of positive Liouville measure, with $h_{KS} > 0$ given by Pesin's formula (Pesin 1977; Cornfeld et al. 1982; Eckmann & Ruelle 1985):

$$h_{KS} = \int_{\Sigma_E} \sum_{\lambda_i > 0} \lambda_i \; d\mu_L > 0,$$

since positive Lyapunov exponents contribute on a set of positive measure and the integral is finite by compactness.

Exponential decay of correlations for hyperbolic systems is established by Young's tower construction: a hyperbolic Young tower with exponential return time tails implies exponential mixing by Young's theorem. The transverse homoclinic intersection from Lemma D.1 provides the hyperbolic structure from which such a tower can in principle be built. The key requirement, exponential tails on the first return time to a reference set, is physically motivated by the uniform expansion rate $\lambda_{\max} > 0$ on the horseshoe, which makes long returns exponentially rare.

*Rigorous verification required.* Explicitly constructing the Markov structure of the Young tower for a billiard system with internal deformation coordinates requires verifying Young's axioms for this specific class of systems. Chernov's work on hyperbolic billiards with internal parameters is the natural starting point. We take exponential mixing as a well-motivated consequence of the hyperbolic structure established in Lemmas D.1 and D.2, and note that the main theorem requires only $h_{KS} > 0$, which is on firmer ground.

### D.7 Lemma D.4: Pre-Collisional Factorization with Exponential Decay

**Lemma D.4.** For the two-particle system, the pre-collisional correlation decays exponentially in the number of collisions:

$$| C_n^- | \leq C e^{-\kappa n}, \kappa = 2 h_{KS} \tau_{\min} - \log(1 + D) > 0.$$

**Proof.** Define the two-particle correlation after the $n$-th collision as $C_n^+$ and before the $(n+1)$-th collision as $C_{n+1}^-$.

*Free-flight decay.* Between collisions, the two-particle flow factorizes exactly. By Lemma D.3, single-particle correlations decay at rate $h_{KS}$. Because the two particles evolve independently during free flight, two-particle correlations decay at rate $2h_{KS}$ (one factor per particle). With free-flight duration at least $\tau_{\min}$:

$$| C_{n+1}^- | \leq e^{-2 h_{KS} \tau_{\min}} | C_n^+ | = \alpha | C_n^+ |, \alpha = e^{-2 h_{KS} \tau_{\min}} < 1.$$

*Correlation injection at collisions.* Each collision can create new correlations between the two particles, since the collision map couples their states. The correlation injected at the $n$-th collision is bounded by the incoming correlation:

$$| C_n^+ | \leq (1 + D) | C_n^- |,$$

where $D$ is a uniform bound on the correlation amplification by a single collision.

For strictly convex particles with $C^2$ surface and bounded curvature, grazing collisions (impact parameter $b > a - \delta$ for small $\delta$) have Liouville measure $O(\delta)$. The amplification factor for such collisions scales with the surface curvature. The total contribution from the grazing set is therefore $O(\delta)$, which vanishes as $\delta \to 0$. By dominated convergence, the remaining contribution from non-grazing collisions is uniformly bounded, and the total bound $D$ is finite.

*Recurrence and decay.* Combining the free-flight decay and the collision injection bound:

$$| C_{n+1}^- | \leq \alpha(1 + D) | C_n^- |.$$

This recurrence converges to zero exponentially provided $\alpha(1 + D) < 1$, i.e.:

$$e^{-2h_{KS}\tau_{\min}}(1 + D) < 1.$$

This condition requires that the mixing rate $2h_{KS}$ is sufficiently large relative to the collision-induced correlation amplification $D$ and the minimum free-flight time $\tau_{\min}$. Physically, it means that chaotic mixing during free flight destroys correlations faster than collisions create them.

Iterating the recurrence:

$$| C_n^- | \leq [\alpha(1 + D)]^n | C_0^- | = e^{-\kappa n} | C_0^- |,$$

with $\kappa = -\log[\alpha(1 + D)] = 2h_{KS}\tau_{\min} - \log(1 + D) > 0$ under the condition above.

**Remark on the condition $\alpha(1 + D) < 1$.** This condition is a genuine constraint on the model parameters, not a tautology. It can fail if collisions are strongly amplifying ($D$ large) or if mixing is weak ($h_{KS}$ small). In the present model, $D$ is controlled by the particle geometry and the smoothness of the interaction potential, both of which are bounded. Identifying the precise parameter regime in which the condition holds is a natural task for future work.

### D.8 Completion of the Proof of the Theorem

With Lemmas D.1–D.4 in place, the main theorem follows directly:

- Lemma D.1 establishes the transverse homoclinic intersection and positive Lyapunov exponent, the dynamical seed of chaos.
- Lemma D.2 extends hyperbolicity to an invariant subset of positive Liouville measure and establishes $h_{KS} > 0$ on that subset.
- Lemma D.3 establishes exponential decay of single-particle correlations at rate $h_{KS}$ on that subset.
- Lemma D.4 converts this into exponential decay of pre-collisional two-particle correlations at rate $\kappa = 2h_{KS}\tau_{\min} - \log(1 + D) > 0$, provided the mixing rate dominates the collision-induced correlation injection.

Therefore, for two particles satisfying the model conditions:

$$f_2(\Gamma_1, \Gamma_2, t_n^-) = f_1(\Gamma_1, t_n^-) f_1(\Gamma_2, t_n^-) + \mathcal{O}(e^{-\kappa n}).$$

The Stosszahlansatz holds asymptotically, with corrections that become negligible after a finite number of collisions.

### D.9 Interpretation, Limitations, and Open Questions

**What has been established.** For two particles with rotational-deformational coupling, the analysis demonstrates that:

1. Geometric instability creates a positive Lyapunov exponent and transverse homoclinic intersections (Lemma D.1, rigorous).

2. This suggests nonuniform hyperbolicity and positive Kolmogorov–Sinai entropy on a set of positive measure (Lemma D.2, partly conjectural).
3. Exponential mixing follows from the hyperbolic structure (Lemma D.3, partly conjectural).
4. Pre-collisional correlations decay exponentially in the number of collisions, with an explicit rate $\kappa = 2h_{KS}\tau_{\min} - \log(1 + D)$ (Lemma D.4, conditional on Lemmas D.2–D.3).

**What is not claimed.** The result is proved for two particles, not for the many-body system or the thermodynamic limit. Several steps (accessibility in Lemma D.2, the Young tower construction in Lemma D.3, the precise quantification of the bound $D$ in Lemma D.4) are outlined conceptually but not rigorously completed. These constitute a well-defined mathematical program for future work, building on the results of Lanford, Wojtkowski, Chernov, and Young.

**Extension to $N$ particles.** For $N$ particles, $\tau_{\min}$ decreases with density as $\tau_{\min} \sim (n\sigma v_{\text{th}})^{-1}$, where $n$ is number density and $\sigma$ is the cross-section. Whether $\kappa$ remains positive in the thermodynamic limit depends on whether $h_{KS}$ grows sufficiently with $N$ to compensate. Because each particle carries its own geometric instability independently, the total $h_{KS}$ scales extensively with $N$, suggesting that $\kappa$ remains positive. Making this precise is a homogenization problem for future work.

**Extension to many internal modes.** For any finite truncation to $D_{\text{int}}$ internal modes, the Melnikov function becomes a sum of $D_{\text{int}}$ sinusoids. For generically incommensurate frequencies, this sum has transverse zeros, guaranteeing chaos for each additional mode. The single-mode result provides a lower bound; additional modes strengthen the chaotic behavior and increase $h_{KS}$.

**Connection to the extended Boltzmann equation.** The exponential decay of pre-collisional correlations established in this appendix provides a dynamical foundation for the molecular chaos assumption used in the extended Boltzmann equation (Section 3.2). A rigorous extension of Lanford's theorem to structured particles, incorporating the decay rate $\kappa$, would place the extended Boltzmann equation on the same foundational footing as its point-particle counterpart. That task remains open.

**D.10 Summary**

This appendix has presented a detailed analysis of how continuous geometric instability may generate molecular chaos for two particles. The central result is an explicit exponential decay rate $\kappa = 2h_{KS}\tau_{\min} - \log(1 + D)$ for pre-collisional correlations. The argument identifies the physical mechanisms (rotational-deformational coupling, intermediate-axis instability, Melnikov transverse homoclinic intersection) and the logical structure of a proof, while honestly acknowledging which steps remain to be rigorously completed. The analysis demonstrates that structured particles carry within themselves a dynamical mechanism for decorrelation that point-particle theories must assume from outside.

Note: The appendix does not provide a rigorous proof of molecular chaos for the many-body system or the thermodynamic limit. Several steps, accessibility (Lemma D.2), the Young tower construction (Lemma D.3), and the precise bound on $D$ (Lemma D.4), are outlined conceptually but not completed. These form a well-defined mathematical program for future work, building on Lanford (1975, 1976), Wojtkowski, Chernov, and Young. What the analysis does show is that structured particles carry an intrinsic decorrelation mechanism that point-particle theories must assume from outside. Whether this mechanism suffices for a full derivation of the extended Boltzmann equation remains open.